\documentclass[
  journal=jacsat,
  manuscript=article,
  oneside
]{achemso}
\usepackage{chemformula} 
\usepackage[T1]{fontenc} 
\usepackage{graphicx}  % needed for figures
\usepackage{amsfonts}
\usepackage{dcolumn}   % needed for some tables
\usepackage{bm}        % for math
\usepackage{amssymb}   % for math
\usepackage[export]{adjustbox}  
\let\oldAA\AA
\renewcommand{\AA}{\text{\normalfont\oldAA}}

\title{Direct nanoscale mapping of band alignment in single-layer semiconducting lateral heterojunctions}

\author{Chakradhar Sahoo}
\affiliation[Aarhus University]
{Department of Physics and Astronomy, Aarhus University, 8000 Aarhus C, Denmark.\\}
\email{sahoo1@phys.au.dk}
\author{Suman Kumar Chakraborty}
\affiliation[IIT Kgp]
{Materials Science Centre, Indian Institute of Technology, Kharagpur 721302, India.\\}
\altaffiliation{These authors contributed equally as second authors\\}
\author{A. Kousika}
\affiliation[IISC]
{Department of Chemical Engineering, Indian Institute of Science, Bengaluru, Karnataka 560012, India.\\}
\altaffiliation{These authors contributed equally as second authors\\}
\author{Alfred J. H. Jones}
\affiliation[Aarhus University]
{Department of Physics and Astronomy, Aarhus University, 8000 Aarhus C, Denmark.\\}
\author{Manas Sharma}
\affiliation[IISC]
{Department of Chemical Engineering, Indian Institute of Science, Bengaluru, Karnataka 560012, India.\\}
\author{Thomas S. Nielsen}
\affiliation[Aarhus University]
{Department of Physics and Astronomy, Aarhus University, 8000 Aarhus C, Denmark.\\}
\author{Zhihao Jiang}
\affiliation[Aarhus University]
{Department of Physics and Astronomy, Aarhus University, 8000 Aarhus C, Denmark.\\}
\author{Ihsan A. Kolasseri}
\affiliation[Aarhus University]
{Department of Physics and Astronomy, Aarhus University, 8000 Aarhus C, Denmark.\\}
\author{Subhadip Das}
\affiliation[Aarhus University]
{Department of Physics and Astronomy, Aarhus University, 8000 Aarhus C, Denmark.\\}
\author{Matthew D. Watson}
\affiliation{Diamond Light Source, Division of Science, Didcot, OX110DE UK.\\}
\author{Cephise Cacho}
\affiliation{Diamond Light Source, Division of Science, Didcot, OX110DE UK.\\}
\author{Kenji~Watanabe}
\affiliation{Research Center for Electronic and Optical Materials, National Institute for Materials Science, 1-1 Namiki, Tsukuba 305-0044, Japan.}
\author{Takashi~Taniguchi}
\affiliation{Research Center for Materials Nanoarchitectonics, National Institute for Materials Science,  1-1 Namiki, Tsukuba 305-0044, Japan.}
\author{Yong P. Chen}
\affiliation[Aarhus University]
{Department of Physics and Astronomy, Aarhus University, 8000 Aarhus C, Denmark.\\}
\author{Tony F. Heinz}
\affiliation{Department of Applied Physics, Stanford University, Stanford, CA 94305, United States.\\}
\author{Ananth Govind Rajan}
\affiliation[IISC]
{Department of Chemical Engineering, Indian Institute of Science, Bengaluru, Karnataka 560012, India.\\}
\author{Prasana K. Sahoo}
\affiliation[IIT Kgp]
{Materials Science Centre, Indian Institute of Technology, Kharagpur 721302, India.\\}
\email{prasana@matsc.iitkgp.ac.in}
\author{Søren Ulstrup}
\affiliation[Aarhus University]
{Department of Physics and Astronomy, Aarhus University, 8000 Aarhus C, Denmark.\\}
\email{ulstrup@phys.au.dk}

\begin{document}

\maketitle

\begin{abstract}
Atomic-scale control over band alignment in single-layer lateral heterostructures (LHSs) of dissimilar transition metal dichalcogenides (TMDCs) is critical for next-generation electronic, optoelectronic, and quantum technologies. However, direct experimental access to interfacial electronic states with nanometer precision remains a significant challenge. Here, we employ angle-resolved photoemission spectroscopy with nanoscale spatial resolution (nanoARPES) to directly map the epitaxial alignment and valence band evolution across MoSe$_{2}$-WSe$_{2}$ LHSs. By combining nanoARPES with spatially resolved photoluminescence, we correlate the evolution of the valence band maximum and exciton features across both atomically sharp and compositionally graded diffusive interfaces. We identified type-II band alignments governed by both material composition and interstitial-induced modifications of band offsets, in close agreement with density functional theory calculations. These results reveal fundamental mechanisms of electronic structure modulation at 1D TMDC heterointerfaces and provide a robust platform for tailored band engineering in van der Waals materials.
 \end{abstract}

%\abbreviations{TMDC, ARPES, PL, LHS}
%\keywords{Nano-ARPES, Electronic structure, lateral heterostructure, band alignment, excitons in TMDC}

Heterostructures composed of two-dimensional (2D) transition metal dichalcogenides (TMDCs) have emerged as versatile platforms for next-generation electronic and optoelectronic devices \cite{liu2016van,castellanos2022van,rivera2015observation,wang2018colloquium}. Their electronic and optical properties are governed by interfacial band alignment, which can be engineered to induce new functionalities \cite{gong2014vertical, huang2014lateral, sahoo2018one, li2015epitaxial, zhang2017robust, yang2017van, li2017two}. Among these systems, monolayer 1D lateral heterostructures (LHSs) are formed by two dissimilar TMDC domains connected through atomically seamless interfaces \cite{sahoo2018one, li2015epitaxial, vashishtha2025epitaxial}. Such junctions minimize contact resistance and thereby facilitate efficient charge transport across the atomically thin boundary \cite{huang2014lateral, gong2014vertical, duan2014lateral, li2015epitaxial}. Furthermore, spatial modulation of chemical composition or intentional alloying across the interface enables the creation of compositionally graded 2D semiconductors, namely diffusive interfaces with tunable band gaps \cite{zheng2018band, zhou2013intrinsic, nugera2022bandgap}. It has been observed that 1D LHSs can experience local strain due to differential thermal expansion during growth and cooling~\cite{kucinski2024direct,zhong2022unified}, promoting the formation of point defects~\cite{gong2014vertical,zhang2018strain,komsa2012two}. Optical spectroscopy of LHSs has revealed band alignment, band gap engineering, and switchable photoresponse across the interface by probing exciton resonances \cite{rosati2023interface, chiu2015determination, berweger2020spatially}. Field-effect modulation has further demonstrated the tunability of excitonic and charged exciton states in these systems \cite{garg2021nanoscale, duan2014lateral, li2017composition, chen2015electronic, zhang2015synthesis, berweger2020spatially}. 

Although optical measurements can access energies of excitonic states and their changes across the LHS interfaces, they do not directly probe the ground-state electronic spectrum to understand the overall band alignment. Scanning tunneling spectroscopy, which measures the momentum-integrated electronic spectrum, has been used to investigate strain-dependent band profiles \cite{zhang2018strain}, band offsets, and edge states in monolayer-to-bilayer transitions \cite{zhang2016visualizing, chu2018energy} within LHS systems. On the other hand, angle-resolved photoemission spectroscopy (ARPES) enables energy- and momentum-resolved electronic structure measurements. To access the momentum-resolved electronic spectrum of TMDC domains in an LHS, nanoARPES is required to directly characterize how interface stoichiometry governs the overall band alignment. This technique has been successfully employed to investigate the direct band gap, indirect-to-direct band gap transitions, quasiparticle band alignment, and defect density-induced electronic structure in single-layer TMDCs \cite{ulstrup2019nanoscale, zhang2014direct, miwa2015van, madeo2020directly, kastl2019effects}. It has also been used to reveal band hybridization, band alignment, moiré superlattices, and flat bands in vertically stacked twisted heterostructures \cite{wilson2017determination, stansbury2021visualizing, gatti2023flat, karni2022structure}.

The above optical and scanning probe studies have provided valuable insights into the spatially dependent band alignment and momentum-integrated band-edge profiles across LHS interfaces. However, direct experimental measurement of the energy- and momentum-resolved variations in the electronic structure across stoichiometry-engineered interfaces remains elusive. In this study, we address these challenges by employing nanoARPES to directly map the electronic structure across monolayer TMDC LHSs. In addition, we use spatially resolved photoluminescence on the micrometer scale ($\mu$PL) across the stoichiometry-engineered interface to probe variations in the excitonic states. We represent the interface structure in our LHSs by the chemical formula W$_{1-x}$Mo$_x$Se$_2$ with $0 \leq x \leq 1$, where the composition $x = 1$ corresponds to MoSe$_{2}$ and $x = 0$ for WSe$_{2}$. We refer to the interface as being "sharp" when the transition from WSe$_{2}$ to MoSe$_{2}$ occurs over an atomic scale with an abrupt change of $x$ from 0 to 1\cite{sahoo2018one}. Conversely, we refer to the interface as "diffusive" when the value of $x$ varies over a length scale of a few nanometers to several micrometers \cite{nugera2022bandgap}. Using a combination of $\mu$PL and nanoARPES spectroscopy, we provide a comprehensive overview of how the electronic structure, optical properties, and overall band alignment are influenced by the stoichiometry at LHS interfaces. Figure 1a shows the stacking geometry of the prepared sample and the nanoARPES experimental configuration, equipped with a zone plate and an order-sorting aperture (OSA).

\begin{figure}
  \includegraphics[width=1\textwidth]{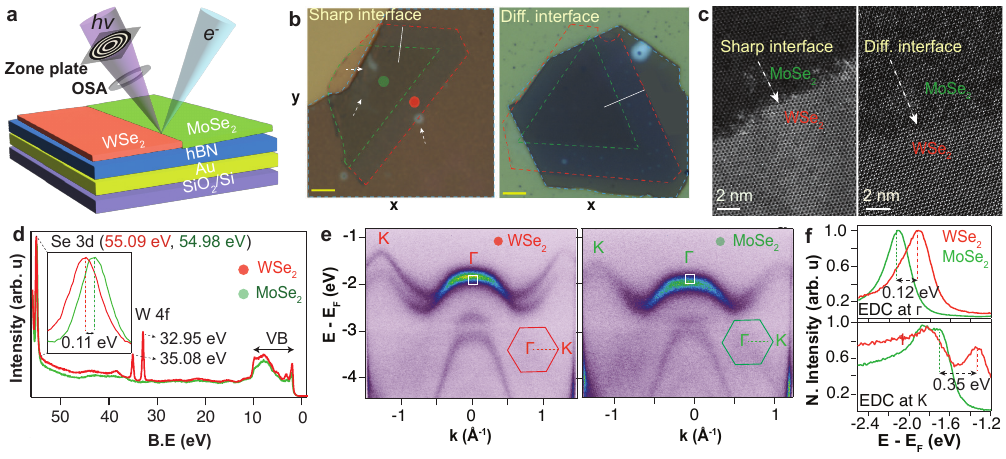}
  \caption{(a) Schematic of the nanoARPES experiment on a monolayer  MoSe$_{2}$-WSe$_{2}$ LHS on hBN atop a gold-coated SiO$_2$/Si substrate. (b) Optical micrographs of the two samples that have the stated interface properties; green and red dotted boundaries are MoSe$_2$ and WSe$_2$ domains respectively. Scale bars are 5~µm unless stated otherwise. (c) Atomic-resolution HAADF-STEM images of both interface types. (d) Momentum-integrated core-level photoemission spectra from WSe$_2$ and MoSe$_2$ domains, the inset shows a magnified view of the Se 3d peaks for MoSe$_2$ and WSe$_2$. (e) Energy- and momentum-resolved ARPES intensity along the $\Gamma$–K direction. Spectra from MoSe$_2$ and WSe$_2$ were obtained from the areas demarcated by green and red dots in (b). (f) Energy distribution curves (EDCs) at $\Gamma$ and K for WSe$_2$ and MoSe$_2$ represented by red and green curves, respectively.}
\label{fig:boat1}
\end{figure}

To investigate the variation in the electronic structure as a function of interface stoichiometry, we prepared MoSe$_{2}$-WSe$_{2}$ LHS samples with both sharp and diffusive interfaces (see optical micrographs of the samples in Figure 1b). The LHS flakes were grown on Si/SiO$_{2}$ substrates using a one-pot, water-assisted CVD strategy \cite{sahoo2018one}. The sharp and diffusive interface LHSs were confirmed by atomic-resolution Z-contrast imaging via high-angle annular dark-field scanning transmission electron microscopy (HAADF-STEM) (see Figure 1c). The LHS flakes were dry-transferred \cite{caldwell2010technique, yi2015review} onto hexagonal boron nitride (hBN) atop a gold-coated Si/SiO$_{2}$ substrate. One part of the LHS flake was kept in contact with the gold layer to ensure proper grounding (see Figure 1b). Prior to the nanoARPES measurements, the samples were annealed overnight in ultrahigh vacuum at a pressure of \( 1 \times 10^{-8} \) mbar at 250\,\textdegree C. 

Before acquiring valence-band (VB) spectra, we performed core-level photoemission spectroscopy to identify the elemental composition across the MoSe$_{2}$-WSe$_{2}$ LHS interfaces. Figure 1d shows the presence of W 4f core-level peaks at a binding energy of $32.95 \pm 0.02$~eV and $35.08 \pm 0.02$~eV in the WSe$_{2}$ domain \cite{lv2022femtomolar}, and their absence in the MoSe$_{2}$ domain. Additionally, a shift of 0.11 eV is observed between the Se 3d peaks of the two materials \cite{kang2018direct}. These spectral features were subsequently used to delineate the material boundaries and interface lengths using W and Se core-level maps \cite{SahooLHSSI}. NanoARPES measurements were performed to image VB spectra along the $\Gamma$–K direction on both domains of the LHS, using identical sample orientation and analyzer geometry. The ARPES (E,k)-spectra (see Figure 1e) show that WSe$_{2}$ and MoSe$_{2}$ share the same in-plane orientation along the $\Gamma$–K direction, confirming epitaxial alignment across the LHS. Energy-distribution curves (EDCs) were extracted at both the $\Gamma$ and K-points (see Figure 1f). At the K points, VBM lies at $-1.32 \pm 0.02$~eV for WSe$_{2}$ and $-1.67 \pm 0.02$~eV for MoSe$_{2}$, giving a VBM offset of $0.35 \pm 0.02$~eV across the junction. We observe two VBM peaks separated by $0.18 \pm 0.03$~eV in MoSe$_{2}$ and $0.48 \pm 0.02$~eV in WSe$_{2}$, consistent with the expected spin-orbit splitting \cite{nakamura2020spin}. At the $\Gamma$-point, the EDCs reveal peaks at $-1.90 \pm 0.02$~eV for WSe$_{2}$ and $-2.02 \pm 0.02$~eV for MoSe$_{2}$, corresponding to a band offset of 0.12 eV. This distinct offset of 0.35 eV at K or 0.12 eV at the $\Gamma$-point raise the question about how the band energies and thus the band alignment evolve across the interface.

\begin{figure}
  \includegraphics[width=1\textwidth]{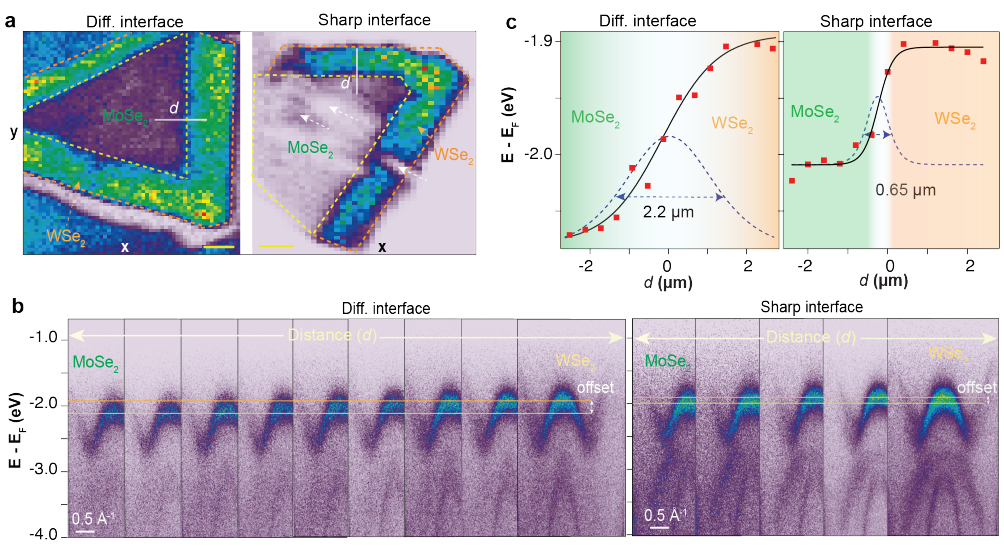}
  \caption{(a) NanoARPES intensity maps of VB near the $\Gamma$-point across diffusive and sharp interface LHS, highlighting the MoSe$_{2}$ and WSe$_{2}$ domains in yellow and orange colors respectively. (b) ARPES (E,k) spectra across the interface, moving from left (MoSe$_{2}$) to right (WSe$_{2}$) side with a step of 0.5 $\mu$m. (c) Line scans (extracted along white line in (a)) at the $\Gamma$-point of VB across both diffusive and sharp interfaces. The red dots represent the extracted peak energies along the scan direction, perpendicular to the interface. The black curves are fitted profiles using sigmoid functions. The widths of both diffusive and sharp interfaces were extracted by differentiating the fit function, as shown as blue dotted curves with their full width at the half maximum (FWHM).}
  \label{Fig2}
\end{figure}

To probe the spatial evolution of the VB energy, we performed four-dimensional nano-ARPES mapping over the energy, momentum, and spatial coordinates $(E, k, x, y)$ on both diffusive and sharp interface LHS samples. This approach enables extraction of the photoemission spectra containing the electronic dispersion $E(k)$ at each $(x, y)$ position on the sample.
Spatial intensity maps were generated (see Figure 2a) by integrating the photoemission signal over an energy window of 0.20~eV and a momentum window of 0.02~\AA$^{-1}$, both centered at the $\Gamma$-point (indicated by the white squares in Figure 1e). This 0.20~eV energy window captures contributions from both WSe$_2$ and MoSe$_2$ VB edges. 
Thus, spatial variations in the $\Gamma$-point VB energy appear as contrast differences between material domains in the $(x, y)$ intensity map. In Figure 2a, the inner domains outlined by yellow dashed lines correspond to MoSe$_2$, while the surrounding regions (highlighted in orange) correspond to WSe$_2$, comparable to the optical image in Figure 1b. The color contrast in the intensity maps likely arises from variations in energy offset across the two domains of each samples. Local inhomogeneity in the $\Gamma$-point energy variations were observed at different regions of the sharp interface (marked by white arrows). This inhomogeneity likely originates from electrostatic variations induced by nanocracks or blisters beneath the hBN, which may introduce local strain during stacking. These features are also visible in the optical image (see Figure 1b), where they appear as regions marked with white arrows. However, all quantitative analyses were restricted to homogeneous interface regions to avoid artifacts arising from cracks or blisters. 

To examine how the band dispersion evolves across the interface, we extracted the local \textit{(E,k)} spectra at successive \textit{(x,y)} positions along the scan lines indicated in Figure 2a (see Figure 2b). The spectra were collected while scanning from the MoSe$_2$ side to the WSe$_2$ side in 0.5~$\mu$m steps for both interface types. As the scan progressed from MoSe$_2$ to WSe$_2$, the $\Gamma$-point VB energy shifted upward, as seen in Figure 2b. To quantify this evolution, we extracted the $\Gamma$-point EDCs as a function of distance $d$ across the interface \cite{SahooLHSSI}. The extracted $\Gamma$-point mean energies are plotted as a function of $d$ in Figure 2c, revealing the spatial evolution of the VBM across the interface. For the sharp interface, the $\Gamma$-point energy shifts from $-1.91$~eV (MoSe$_2$ side) to $-2.01$~eV (WSe$_2$ side), corresponding to a band offset of 0.11~eV. For the diffusive interface, the $\Gamma$-point energy evolves more gradually from -1.91 eV to -2.08 eV, giving a larger offset of 0.17~eV. By fitting the spatial energy profiles \cite{SahooLHSSI}, we extract interface widths of 2.2~$\mu$m for the diffusive interface and 0.65~$\mu$m for the sharp interface. The latter is limited by the nanoARPES beam-spot size. The change in VBM offset between the diffusive and sharp interfaces is somewhat surprising being the same two materials in both the systems. We extracted the band offsets beyond 2~$\mu$m from the interface and it has been observed that both the sharp and diffusive interface maintain the same band offset away from the interface \cite{SahooLHSSI}. This analysis resolves the spatial variation of the VBM across the LHS and provides a direct picture of the band-edge evolution across the lateral transition region. However, understanding of conduction band or excited state evolution is critical to understand the overall band alignment across these LHS interfaces.

\begin{figure}
  \includegraphics[width=1\textwidth]{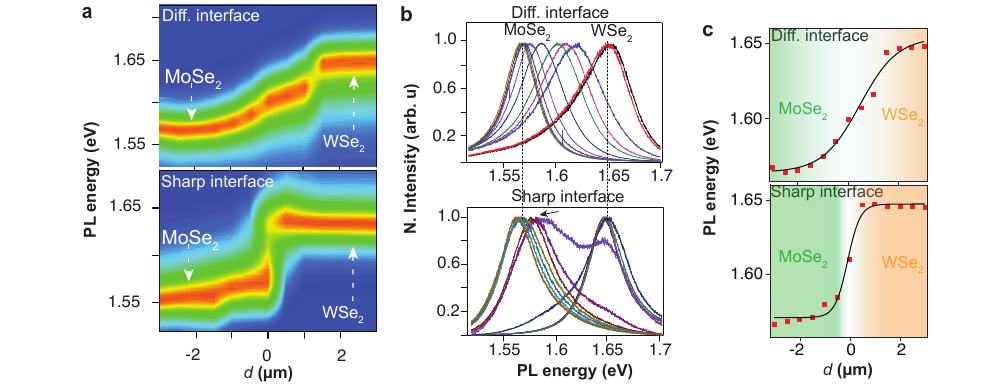}
  \caption{ (a) Contour colour plots of the PL intensity, showing measurement position versus PL energy across the interface for both sharp and diffusive interfaces. (b) PL spectra recorded while scanning laterally from  MoSe$_{2}$ to WSe$_{2}$ across the interface, highlighting spectral shifts near the junction. (c) Extracted peak energy (red dots) of the spectra in (b), the black curves are fitted profiles using sigmoid functions.}
\label{Fig4}
\end{figure}

To investigate the variation in exciton resonances across the heterointerfaces, we performed \(\mu\)PL measurements on both the sharp and diffusive interface samples. Figure 4a shows intensity profile of PL spectra across the heterojunctions, along the white lines indicated in Figure 1b. Figure 4b shows the PL spectra across the interfaces while spatially measuring from MoSe$_{2}$ to WSe$_{2}$ side in steps of 0.5~$\mu$m for both interfaces. On the WSe$_{2}$ side of the LHS, we observe a peak centered at 1.65 eV, which corresponds to the A-exciton of WSe$_{2}$. On the MoSe$_{2}$ side, we see a peak centered at 1.58 eV, which corresponds to the A exciton of MoSe$_{2}$ \cite{sahoo2018one}. In the diffusive interface we see the peak shift between these two values continuously. In contrast, for the sharp interface, the exciton peak transition occurs either at 1.58 eV or 1.65 eV, with a blue shift of 10-12 meV in the MoSe$_{2}$ side (see marked black arrow). The positions along the scan exhibit PL spectra corresponding to either MoSe$_{2}$ or WSe$_{2}$, with a dual-peak spectrum observed only at the interface. This behavior corroborates the presence of a sharply defined boundary \cite{sahoo2018one} of Se based junctions. From the spatially resolved PL spectra and its linear behaviour of the change in PL energy in diffusive interface, the material composition was inferred using a fitting approach \cite{nugera2022bandgap} (see the Supporting Information). This PL study provides an overview of the interface length and hence composition-dependent variations of excitonic states. However, understanding the microscopic origin of these composition-dependent VBM and excitonic states remains essential.

\begin{figure}
 \includegraphics[width=1.0\textwidth]{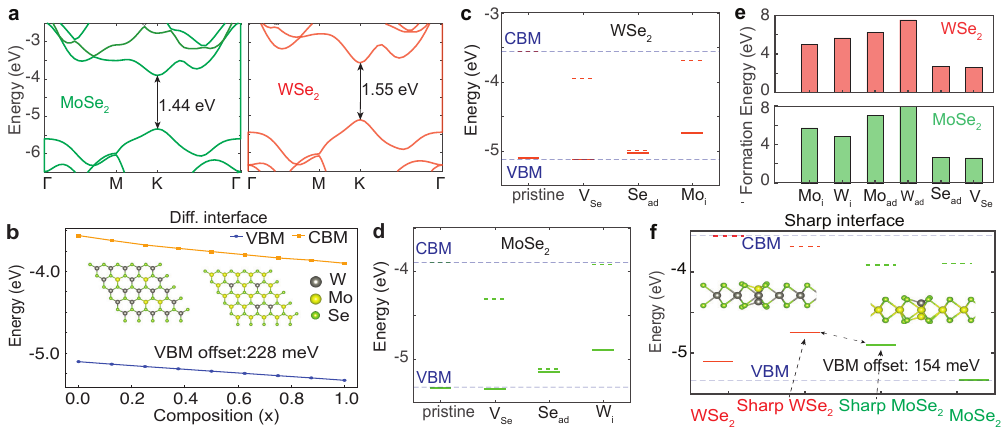}
\caption{(a) Calculated band structure of pristine MoSe$_2$ and WSe$_2$. (b) VBM and conduction band minimum (CBM) energies of W$_{1-x}$Mo$_x$Se$_2$ alloys ($0 < x < 1$) with varying composition. Inset: Top-view of relaxed structures for W-rich ($x = 0.25$) and Mo-rich ($x = 0.75$) alloys; gray, yellow, and green denote W, Mo, and Se atoms, respectively.  (c–d) Energy level diagrams showing defect-induced band edges in (c) WSe$_2$ and (d) MoSe$_2$. Red dashed lines represent VBM/CBM of pristine materials, green solid/dashed lines show occupied/unoccupied defect states. (e) Formation energies of different defects: Mo interstitial (Mo$_i$), W interstitial (W$_i$), Mo adatom (Mo$_{ad}$), W adatom (W$_{ad}$), Se adatom (Se$_{ad}$), and Se vacancy (V$_{Se}$). (f) Energy level diagram for the MoSe$_2$/WSe$_2$ sharp interface. Red line: WSe$_2$; green lines: MoSe$_2$ solid lines: VB; dashed lines: CB. Black dashed double-sided arrow marks the VBM offset. Inset: Side-view of Mo interstitial in WSe$_2$ and W interstitial in MoSe$_2$, used as models for the interfacial regions.}
  \label{fig:boat1}
\end{figure}

It is notable that the overall band offset at the sharp interface is lower than at the diffusive interface, though the offset per unit length is greater for the sharp case. To understand the evolution of band energies and its microscopic origin across the heterojunction, we performed quantum-mechanical density functional theory (DFT) calculations. Figure~4a shows the band structures of pristine MoSe$_2$ and WSe$_2$. The diffusive interface was modeled as a compositionally varying alloy W$_{1-x}$Mo$_x$Se$_2$ ($0 < x < 1$), in line with experimental observations and the natural tendency for alloying at the interface~\cite{nugera2022bandgap,gong2014vertical}. As shown in Figure~4b, both VB and CB edges decrease smoothly with increasing Mo content, indicating a gradual electronic transition. The VB offset and CB offset between pristine WSe$_2$ and MoSe$_2$ are approximately 0.23\,eV and 0.34\,eV, respectively.

To explore the reason for the experimentally observed reduced VB offset at the sharp MoSe$_2$/WSe$_2$ LHS, we considered factors such as local strain and point defects, which are known to alter orbital overlap and band dispersion~\cite{zhang2018strain,nugera2022bandgap,shaposhnikov2019impact}. Interstitials, with low migration barriers (0.1--0.5~eV), tend to migrate toward sharp interfaces and accumulate there due to strain-induced trapping~\cite{komsa2015native,han2018strain}. In contrast, diffusive interfaces distribute strain more uniformly, reducing defect formation~\cite{bogaert2016diffusion,taghinejad2018defect}.

To assess defect-driven band alignment, we investigated the effects of Se vacancies, adatoms, and interstitials on MoSe$_2$ and WSe$_2$ electronic structure using a 6$\times$6$\times$1 supercell and a $\Gamma$-centered 2$\times$2$\times$1 k-point mesh. Optimized structures are shown in the Supporting Information \cite{SahooLHSSI}. Se adatoms and vacancies exhibit the lowest formation energies (Figure~4e), consistent with previous studies~\cite{haldar2015systematic,holtzman2024equilibrium}. These defects introduce states near the pristine VB edge, minimally affecting the band structure of WSe$_2$ and MoSe$_2$ (Figures~4c, 4d). In contrast, defects involving transition metal atoms, such as Mo or W interstitials, introduce mid-gap states positioned approximately 0.2--0.5~eV away from the pristine VB edge. These states can significantly alter the position of the VB maximum and, consequently, the band gap. While Se-related defects are more thermodynamically favorable, metal interstitials may form at high synthesis temperatures (1000--1200~K) due to entropic effects. We explicitly modeled presence of interstitial point defects to evaluate their impact on band alignment (Figure~2). Based on this, we modeled the sharp interface using MoSe$_2$ with a W interstitial and WSe$_2$ with a Mo interstitial (Figure~4f). The VB offset and CB offset between these defective structures were calculated as 0.15\,eV and 0.23\,eV, respectively, consistent with the experimentally observed reduction in the VBM offset. These results suggest that point defects at the sharp interface may account for the lower band offset compared to the diffusive interface.

  \begin{figure}
  \includegraphics[width=0.5\textwidth]{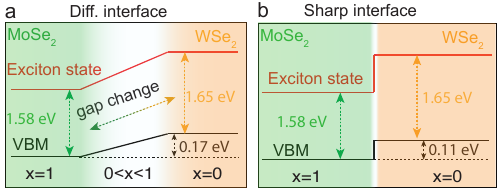}
  \caption{Schematic of composition dependent band alignment across the diffusive and sharp interface, where black and red lines represent the VBM and exciton states respectively. (a) Interplay between band offset, band alignment, and excitonic states across diffusive interface, and (b) interplay between band offset, band alignment, and excitonic states across sharp interface.}
 \label{Fig6}
\end{figure}

From the nanoARPES and PL studies, we visualize the composition dependent band alignment through a band diagram (see the schematic in Figure 5a, 5b) across the LHS interfaces. For the diffusive interface, the VBM offset of 0.17 eV is observed between MoSe$_{2}$ and WSe$_{2}$ accompanied by a continuous change in VBM energy with composition over the range $0 < x < 1$. A corresponding continuous shift in PL energy from 1.58 eV to 1.65 eV is also observed within the same composition range. This observation indicates the graded alloying or chemical intermixing at the interface\cite{feng2014growth, duan2016synthesis}. In contrast, for the sharp interface, both PL and VBM transitions occur within the spatial resolution of the measurement spot, exhibiting a VBM offset of 0.11 eV. The band diagram includes the variation of both the energetic shift in VBM edges, band offsets, and the exciton peak changes (band gap changes) across the interfaces. Our work is in line with recent experiments that demonstrated the composition-dependent band offset changes in Se-based LHSs \cite{nugera2022bandgap, zheng2018band}, measured using kelvin probe microscopy. Our study provides further insights into how the composition influences the band structure and band alignment (Figure 5).  Overall, these composition-driven variations in VBM offset, exciton energies collectively modify the band alignments, highlighting stoichiometric engineering as an effective route to tailor the band structure across 1D LHSs. We note that the exact conduction band position cannot be inferred directly from ARPES. In principle, measurement of the exciton binding energy by PL excitation spectroscopy and nanoARPES measurements on an LHS functional device with the possibility of occupying the conduction band via gating could provide further experimental insights into the interplay of band-renormalization effects \cite{yao2017optically, sahoo2025quasiparticle}.

In summary, nanoARPES combined with spatially resolved photoluminescence provides direct access to the nanoscale electronic structure of monolayer MoSe$_{2}$-WSe$_{2}$ lateral heterostructures. By comparing atomically sharp and compositionally graded interfaces, we show that local stoichiometry and interstitial configurations strongly modulate the valence-band offset and excitonic energies, leading to distinct band-alignment profiles across the junction. These results establish that band alignment in 1D LHSs is not solely determined by the parent materials but is highly sensitive to nanoscale chemical variation. Such insights provide fundamental design rules for engineering band offsets, depletion regions, and charge-separation pathways in LHS-based electronic, optoelectronic, and quantum devices.

 \section{Methods}

\subsection{Crystal Growth} 

To synthesize single-junction 1L MoSe$_{2}$-WSe$_{2}$ LHS, a one-step water-assisted CVD process was employed with a mixture of MoSe$_{2}$ and WSe$_{2}$ bulk precursors, as per the earlier report \cite{sahoo2018one}. The selective growth of individual TMDC layers was realized by switching carrier gases. Bulk precursors were kept at  \(1050^\circ\mathrm{C}\), whereas growth substrates were kept at  \(800^\circ\mathrm{C}\). In the initial growth stage, the N$_2$ + H$_2$O environment facilitates the oxidation of MoSe$_2$ and WSe$_2$ into higher sub-oxides of Mo and W, respectively. The higher volatility of Mo-based sub-oxides promotes the growth of MoSe$_2$. Upon switching the carrier gas to Ar + H$_2$ (5\%), the growth of MoSe$_2$ is terminated due to the rapid reduction of Mo-based sub-oxides to metallic Mo. In contrast, the slower reduction rate of W-based sub-oxides, together with their volatility, provides the adatoms necessary for WSe$_2$ growth at the edges of the pre-grown MoSe$_2$ template. The spatial width, shape, size, and degree of alloying within individual domains are controlled independently.

\subsection*{Sample Preparation}

The heterostructure samples were prepared using a dry transfer stacking method. A polycarbonate (PC) film mounted on a polydimethylsiloxane (PDMS) stamp was employed to pick up the CVD-grown LHS flakes. These flakes were carefully aligned and transferred onto an exfoliated hBN layer. The resulting LHS/hBN stack was subsequently placed onto a gold-coated substrate such that a portion of the LHS flake remained in contact with the gold layer, serving as a grounding point during measurements. The gold coating (thickness of 50~nm) was patterned on a \(\mathrm{Si/SiO_2}\) substrate via photolithography prior to the sample transfer. Post-transfer, the assembled stack resembled the configuration shown in Figure~1a. The samples were cleaned sequentially using chloroform, acetone, and isopropanol, each for 1~minute. Following solvent cleaning, the samples were annealed at \(280^\circ\mathrm{C}\) for 5~hours in a mixed atmosphere of 5\% \(\mathrm{H_2}\) in \(\mathrm{Ar}\), ensuring removal of residual contaminants and enhancing interfacial adhesion \cite{sahoo2025quasiparticle}.

\subsection{Photoemission Measurements} 

The nanoARPES measurements were carried out at the I05 beamline of the Diamond Light Source (DLS), UK. A spatial resolution of approximately 850~nm was achieved using a zone plate focusing configuration. The measurements were performed with a photon energy of 85~eV, and the incident light had predominantly linear horizontal polarization. Photoemitted electrons from the illuminated region are collected by a hemispherical analyzer equipped with a 2D detector, where the sample was oriented along the $\Gamma$-K direction of the Brillouin zone. During the experiment, the sample was maintained at a temperature of $T = 50 $~K. All raw ARPES spectra were transformed from angular to momentum space, and Fermi-level alignment was performed using reference spectra acquired from a gold substrate under identical experimental conditions. All data visualization and analyses were performed using \textit{WaveMetrics IGOR Pro 7}.

\subsection*{Photoluminescence Measurements}

Photoluminescence (PL) measurements were carried out using a commercial Raman-PL spectroscopy system (RENISHAW). A continuous-wave (CW) laser with a wavelength of 532~nm (2.33~eV) was employed as the excitation source. The excitation beam was focused onto the sample surface using a 100$\times$ objective lens with a numerical aperture (N.A.) of 0.85. The same objective was used to collect the emitted PL signal in a back-reflection geometry. The collected light was dispersed by a spectrometer equipped with a 1200~lines/mm diffraction grating and directed onto a 2D charge-coupled device (CCD) detector array. The overall spectral resolution of the system was approximately 300~$\mu$eV. All the PL data were measured at room temperature and analyzed using \textit{Origin 8.5}.

\subsection{Density Functional Theory Calculations} 

 Density functional theory (DFT) calculations were performed on monolayers of WSe$_2$, MoSe$_2$, and alloyed W$_{1-x}$Mo$_x$Se$_2$ ($0 < x < 1$) using the Vienna Ab initio Simulation Package (VASP) 6.4.1 with a plane-wave basis set~\cite{kresse1996efficiency, kresse1994theory}. Ion--electron interactions were described using the projector augmented wave (PAW) method~\cite{kresse1999ultrasoft, blochl1994projector} with a 450~eV energy cutoff. Exchange-correlation effects were treated using the Perdew--Burke--Ernzerhof (PBE) functional~\cite{perdew1996generalized}. A 4$\times$4$\times$1 supercell and a $\Gamma$-centered 2$\times$2$\times$1 k-point mesh~\cite{monkhorst1976special} were used, ensuring convergence within 1~meV/atom. A vacuum spacing of 20~\AA{} was applied along the $c$-axis to suppress interlayer interactions. Structures were optimized using the conjugate gradient method with convergence thresholds of 10$^{-6}$~eV for energy and 10$^{-2}$~eV/\AA{} for forces. Gaussian smearing (0.03~eV) was used to aid convergence near the Fermi level. Lattice constants and bandgaps were computed and validated against experimental data~\cite{schutte1987crystal,zhang2014direct, desai2014strain, lu2017identifying} (see the Supporting Information \cite{SahooLHSSI}). For pristine monolayer unit cells, a denser $\Gamma$-centered 12$\times$12$\times$1 mesh was employed. Although PBE underestimates the absolute experimental band gap, it remains a practical choice for exploring the effect of defects in large supercells. For quantitative agreement with experimental optical gaps, higher-level methods such as hybrid functional plus spin orbit coupling (HSE06+SOC) and GW+BSE would be necessary \cite{sadhukhan2022generating} which are computationally prohibitive for large supercells considered in this study. PBE reliably captures qualitative features such as defect-induced band gap reduction and the appearance of in-gap states in TMDs and TMD based alloys like W$_{1-x}$Mo$_x$Se$_2$ \cite{nakamura2020spin, kresse1996efficiency, kresse1999ultrasoft}, which are the focus of this study.

\section{Acknowledgements}
C.S. acknowledges Marie Sklodowska-Curie Postdoctoral Fellowship (proposal No.101059528). A.K. thanks the Anusandhan National Research Foundation for the National Post-Doctoral Fellowship (Grant No. PDF/2023/000864). P.K.S. acknowledges the Department of Science and Technology (DST) (Project Code: DST/NM/TUE/QM-1/2019 and DST/TDT/AMT/ 2021/003 (G) and (C)) and National Quantum Mission (NQM) project DST/QTC/NQM/ QMD/2024/4/(G), India. TEM work was performed at Sophisticated Analytical Technical Help Institute (SATHI), IIT Kharagpur, supported by DST, Govt of India. The work was funded/co-funded by the European Union (ERC grant EXCITE with Project Number 101124619). Views and opinions expressed are however those of the author(s) only and do not necessarily reflect those of the European Union or the European Research Council. Neither the European Union or the granting authority can be held responsible for them. The authors acknowledge funding from the Novo Nordisk Foundation (Project Grant NNF22OC0079960), VILLUM FONDEN under the Villum Investigator Program (Grant No. 25931). K.W. and T.T. acknowledge support from the JSPS KAKENHI (Grant Numbers 21H05233 and 23H02052) and the World Premier International Research Center Initiative (WPI), MEXT, Japan. A.G.R. acknowledges the use of the Param Pravega machine at the Supercomputer Education and Research Centre, Indian Institute of Science, the Param Yukti machine at the Jawaharlal Nehru Centre for Advanced Scientific Research, and the Param Smriti machine at NABI Mohali.

\section{Supporting Information} 

\section{Details on the CVD growth of LHS flakes} 
The sharp and diffusive interfaces were grown by controlling the switching between N$_2$+H$_2$O and 5 \% of Ar+H$_2$. During switching, an intentional mixing of these two gases promotes co-evaporation of Mo and W-based complexes, resulting in a diffusive interface. On the other hand, an abrupt change/switching avoids the possibility of mixing, turning into a sharp interface. 

Optical images of the chemical vapor deposition (CVD)-grown LHS flakes were captured using a Nikon optical microscope with a 100$\times$ objective lens (numerical aperture = 0.9). In Figure~S1, the central triangular regions correspond to monolayer MoSe$_2$ domains, while the surrounding areas are identified as WSe$_2$. Each image includes a scale bar of 5~$\mu$m. The sharp and diffusive interfaces are marked with blue arrows, with a larger contrast in the case of the sharp interface than the diffusive one. These contrasting interfacial features highlight differences in domain growth kinetics during the CVD growth process. 

\begin{figure}
  \includegraphics[width=0.5\textwidth]{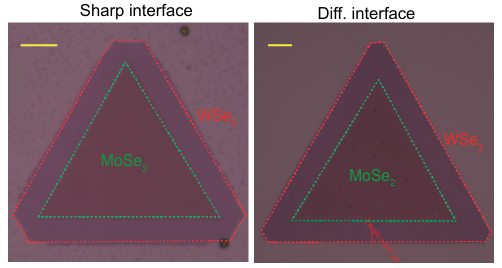}
  \caption{Optical microscopy of MoSe$_2$-WSe$_2$ LHS exhibiting both sharp and diffusive interfaces, where the interface is marked in green.}
  \label{fig:boat1}
\end{figure}

\section{Extraction of EDC, VBM mean energy, and interface width of LHSs}
The energy distribution curves (EDCs) for both sharp and diffusive interfaces are shown in Figure~S2, where the red curves are the nanoARPES data across the $\Gamma$-point and black curves are Lorentzian fits, whose peak energies are plotted in Figure~2 of the main draft. 

\begin{figure}
  \includegraphics[width=0.5\textwidth]{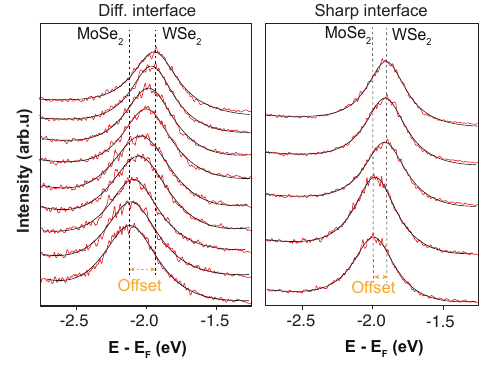}
  \caption{EDCs at the $\Gamma$-point measured across sharp and diffusive interfaces}
  \label{fig:boat1}
\end{figure}

The analysis of the $\Gamma$-point was considered over the K-point as the photoemission intensity was better than at the K-point. The EDCs are plotted at the steps of 0.5 $\mu$m across the interface, whose ARPES (E,k) spectra are shown in Figure~2 of the main draft. 
The valence band maximum (VBM) energies at the sharp and diffusive interfaces of the LHS flake were determined by performing a line scan across the interface while measuring the EDCs at the $\Gamma$- and $K$-points, as shown in Figure~1 and~2 in the main text. The VBM positions were extracted by fitting the EDCs using Lorentzian functions. The extracted energy positions and associated uncertainties across the interfaces are presented in the main draft. These mean VBM energies across the interface were fitted through a sigmoid function, consisting of a step function convoluted with a Lorentzian function. The widths of both diffusive and sharp interfaces were extracted by differentiating the fit function whose full width at half maximum (FWHM) defines the interface width as measured by nanoARPES (see Figure~2 in the main draft).

\section{Extraction of PL mean energies and material composition across the interface} 

\begin{figure}
  \includegraphics[width=1.0\textwidth]{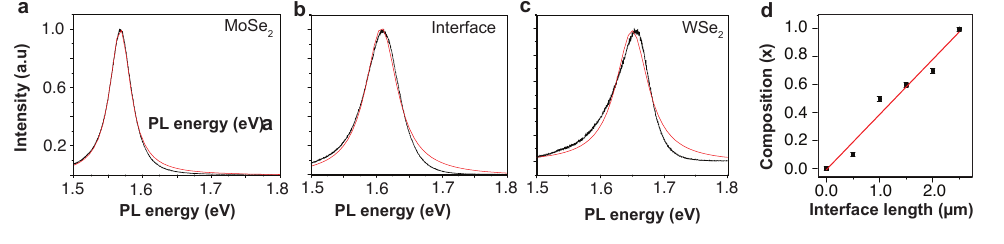}
  \caption{Measured PL spectra on (a) MoSe$_2$, (b) interface, and (c) WSe$_2$ domains of the diffusive interface. The black curves represent the experimental data, while the red curves correspond to the fitted spectra. (d) Extracted compositional values from the mean PL peak energies across the diffusive interface. Black dots are the extracted PL peak energies and the red line is their linear fit.}
  \label{fig:boat1}
\end{figure}
The mean energies of the photoluminescence (PL) spectra were extracted at both sharp and diffusive interfaces of the LHS flake by performing single Gaussian fitting to each spectrum. Figure~S4 presents representative spectra corresponding to WSe$_2$, MoSe$_2$, and at the intermediate region of the MoSe$_2$-WSe$_2$ diffusive interface. The excitonic peaks are centered approximately at 1.58\,eV for WSe$_2$, 1.65\,eV for MoSe$_2$, and between 1.58\,eV and 1.65\,eV for the interfacial region. Based on the extracted mean energies, the local material composition was estimated using Vegard’s law, as illustrated in Figure~S3.

\section{VB offsets away from the interface}
Band offsets are extracted away from the interface as shown in Figure~S4, providing similar values for both sharp and diffusive interfaces. Additionally, within a fixed spatial window (marked by a rectangle), the band offset at the sharp interface is larger than that at the diffusive interface.

\begin{figure}
  \includegraphics[width=0.5\textwidth]{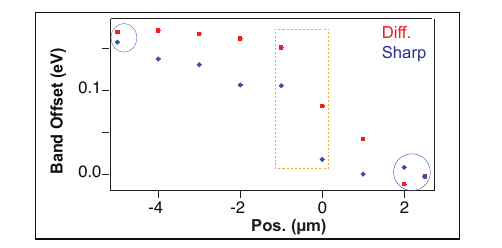}
  \caption{Band offset values of sharp and diffusive interfaces, extracted away from the interface. The offset values are referenced with respect to the WSe$_2$ side of the flakes. The marked rectangle is the band offset within 1 $\mu$m at the iinterface.}
  \label{fig:boat1}
\end{figure}

\begin{figure}
  \includegraphics[width=1\textwidth]{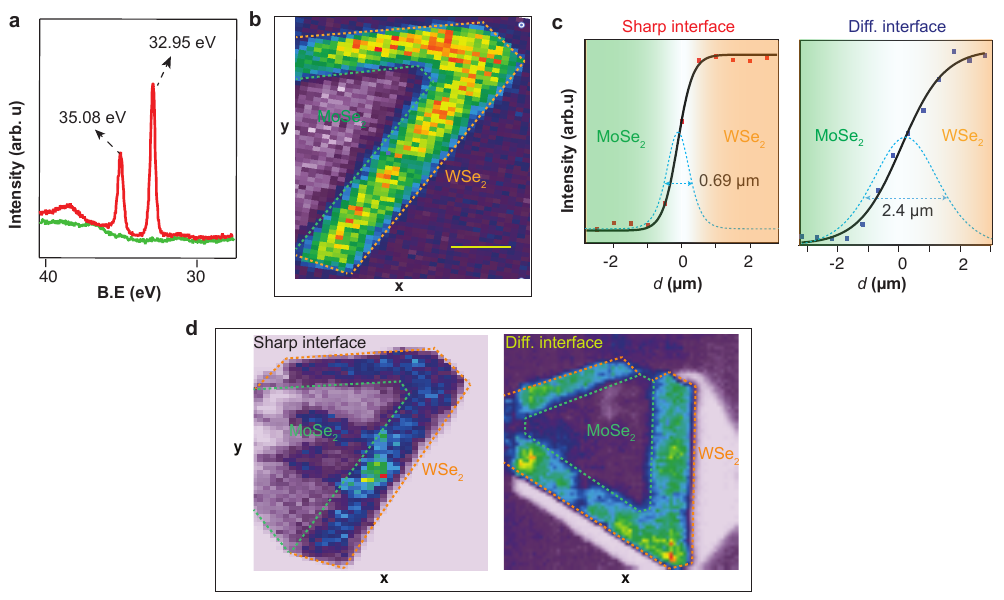}
  \caption{(a) Momentum-integrated core-level photoemission spectra from WSe$_2$ and MoSe$_2$ domains, marked in red and green respectively. (b) W core-level intensity map of the sharp interface, with distinct regions corresponding to WSe$_{2}$ and MoSe$_{2}$, as marked in dotted lines. (c) Line profiles extracted along the interface regions as marked white lines in Figure 1b for sharp and diffusive interfaces. (d) Se core-level intensity map of the sharp interface, with distinct regions corresponding to WSe$_{2}$ and MoSe$_{2}$ marked for both sharp and diffusive interfaces.}
  \label{FigS5}
\end{figure}
\section{Core level study} 
Spatially resolved W core-level measurements were performed to detect the variation of W photoemission intensity across the interfaces (see Figure~\ref{FigS5}a). W core-level intensity map vs (x,y) position (see Figure~\ref{FigS5}b) was generated by integrating the spectral signal over an energy window of 0.8~eV centered at 32.95 eV (see Figure~\ref{FigS5}a). The regions of the outer and inner domains correspond to the WSe$_{2}$ and MoSe$_{2}$ domains as marked in Figure~\ref{FigS5}b. Figure~\ref{FigS5}c shows the corresponding line scans performed across the interface. From the line scan analysis, the interface lengths were determined to be 0.69 $\mu$m for the sharp and 2.4 $\mu$m for diffusive interface. The Se core-level maps for both the sharp and diffusive interfaces are displayed in Figure~\ref{FigS5}d, prepared by binning 0.2 eV of energy at the binding energy of 55 eV. The core level analysis identifies the material boundaries and quantifies the interface lengths.  

\section{Details of the DFT calculations for pristine and defective 2D $M$Se$_2$} 
\begin{table}[h]
\centering
\caption{Comparison of the in-plane lattice constants $a$ (in \AA) and band gaps (in eV) for $M$Se$_2$ ($M$ = Mo, W) as calculated using DFT-PBE, along with corresponding experimental values.}
\begin{tabular}{|c|c|c|}
\hline
\textbf{Parameters} & \textbf{MoSe$_2$} & \textbf{WSe$_2$} \\
\hline
Lattice parameter (PBE) & 3.28 & 3.28 \\
Lattice parameter (Exp.) & 3.29 & 3.28 \\
Band gap (eV) (PBE) & 1.44 & 1.55 \\
Band gap (eV) (Exp.) & 1.58 & 1.65 \\
\hline
\end{tabular}
\end{table}

The electronic band structures and corresponding band gaps were calculated at the PBE-GGA level using non-self-consistent single-point calculations on the relaxed structures. The band structure was computed along the $\Gamma$--M--K--$\Gamma$ path, with 20 $k$-points sampled along each high-symmetry direction in the reciprocal space. The electrostatic potential in the vacuum region was used as the reference energy in each case, and the band energies were aligned accordingly by shifting them relative to this vacuum level.

\begin{figure}
  \includegraphics[width=0.5\textwidth]{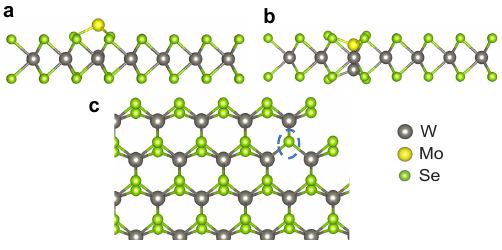}
  \caption{Optimized structures for various defects in the $6 \times 6 \times 1$ supercell of WSe$_2$: (a) Mo adatom, (b) Mo interstitial, and (c) Se vacancy,  indicated by a blue dashed circle. Gray, yellow, and green colors represent W, Mo, and Se atoms, respectively.}
  \label{fig:boat1}
\end{figure}

As a first step, we computed the formation energies of various intrinsic defects, including Mo, W, and Se vacancies, adatoms, and interstitials, to identify the most energetically favorable defect types in $M$Se$_2$. The formation energy $E_f$ of a defect was calculated using the following equation:

\[
E_f = E_{\text{defect}} - E_{\text{pristine}} \pm n\mu_i
\]

where $E_{\text{defect}}$ is the total energy of the supercell containing the defect, $E_{\text{pristine}}$ is the total energy of the pristine supercell, $n$ is the number of atoms removed or added, and $\mu_i$ is the chemical potential of the added or removed atom $i$, depending on whether the defect is an interstitial ($-$) or a vacancy ($+$). The chemical potential of the metal ($M$) was taken as the total energy per atom of its bulk body-centered cubic (bcc) phase. Similarly, the chemical potential of Se was taken as the total energy per atom of the most stable crystalline phase (trigonal) of bulk Se. These reference chemical potentials represent the elemental reservoirs and define the upper bounds in the thermodynamic stability range of the compound.

\section{Author Contribution}
S.K.C. and A.K. contributed equally as second authors. C.S., S.U. and P.K.S. conceived the project. S.K.C. and P.K.S. performed the CVD growth and STEM measurements of the LHS flakes. C.S. prepared the samples using the dry-transfer method. C.S. carried out the PL measurements with assistance from I.A.K. and S.D., and analyzed the PL data together with S.K.C. NanoARPES and core-level measurements were performed by C.S., A.J.H.J., T.S.N., Z.J., S.U., M.D.W. and C.C, where C.S. analyzed the nanoARPES, and core-level data. C.S. wrote the manuscript with inputs from S.U., and P.K.S. A.K. and M.S. performed the DFT calculations with inputs from A.G.R., and together wrote the theoretical sections with A.G.R. All authors provided feedback and comments to the preparation of the final draft of the manuscript.

\section{Conflict of Interest}
The authors declare no conflict of interest.

\section{Data Availability}
The data used in this study is available on the "Zenodo" platform (10.5281/zenodo.1234567).

\bibliography{LHS.bib}

@article{chiu2015determination,
  title={Determination of band alignment in the single-layer MoS2/WSe2 heterojunction},
  author={Chiu, Ming-Hui and Zhang, Chendong and Shiu, Hung-Wei and Chuu, Chih-Piao and Chen, Chang-Hsiao and Chang, Chih-Yuan S and Chen, Chia-Hao and Chou, Mei-Yin and Shih, Chih-Kang and Li, Lain-Jong},
  journal={Nature communications},
  volume={6},
  number={1},
  pages={7666},
  year={2015},
  publisher={Nature Publishing Group UK London}
}

@article{wilson2017determination,
  title={Determination of band offsets, hybridization, and exciton binding in 2D semiconductor heterostructures},
  author={Wilson, Neil R and Nguyen, Paul V and Seyler, Kyle and Rivera, Pasqual and Marsden, Alexander J and Laker, Zachary PL and Constantinescu, Gabriel C and Kandyba, Viktor and Barinov, Alexei and Hine, Nicholas DM and others},
  journal={Science advances},
  volume={3},
  number={2},
  pages={e1601832},
  year={2017},
  publisher={American Association for the Advancement of Science}
}

@article{huang2014lateral,
  title={Lateral heterojunctions within monolayer MoSe 2--WSe 2 semiconductors},
  author={Huang, Chunming and Wu, Sanfeng and Sanchez, Ana M and Peters, Jonathan JP and Beanland, Richard and Ross, Jason S and Rivera, Pasqual and Yao, Wang and Cobden, David H and Xu, Xiaodong},
  journal={Nature materials},
  volume={13},
  number={12},
  pages={1096--1101},
  year={2014},
  publisher={Nature Publishing Group UK London}
}

@article{gong2014vertical,
  title={Vertical and in-plane heterostructures from WS 2/MoS 2 monolayers},
  author={Gong, Yongji and Lin, Junhao and Wang, Xingli and Shi, Gang and Lei, Sidong and Lin, Zhong and Zou, Xiaolong and Ye, Gonglan and Vajtai, Robert and Yakobson, Boris I and others},
  journal={Nature materials},
  volume={13},
  number={12},
  pages={1135--1142},
  year={2014},
  publisher={Nature Publishing Group UK London}
}

@article{duan2014lateral,
  title={Lateral epitaxial growth of two-dimensional layered semiconductor heterojunctions},
  author={Duan, Xidong and Wang, Chen and Shaw, Jonathan C and Cheng, Rui and Chen, Yu and Li, Honglai and Wu, Xueping and Tang, Ying and Zhang, Qinling and Pan, Anlian and others},
  journal={Nature nanotechnology},
  volume={9},
  number={12},
  pages={1024--1030},
  year={2014},
  publisher={Nature Publishing Group UK London}
}

@article{zhang2018strain,
  title={Strain distributions and their influence on electronic structures of WSe2--MoS2 laterally strained heterojunctions},
  author={Zhang, Chendong and Li, Ming-Yang and Tersoff, Jerry and Han, Yimo and Su, Yushan and Li, Lain-Jong and Muller, David A and Shih, Chih-Kang},
  journal={Nature nanotechnology},
  volume={13},
  number={2},
  pages={152--158},
  year={2018},
  publisher={Nature Publishing Group UK London}
}

@article{garg2021nanoscale,
  title={Nanoscale Raman characterization of a 2D semiconductor lateral heterostructure interface},
  author={Garg, Sourav and Fix, J Pierce and Krayev, Andrey V and Flanery, Connor and Colgrove, Michael and Sulkanen, Audrey R and Wang, Minyuan and Liu, Gang-Yu and Borys, Nicholas J and Kung, Patrick},
  journal={ACS nano},
  volume={16},
  number={1},
  pages={340--350},
  year={2021},
  publisher={ACS Publications}
}

@article{li2017composition,
  title={Composition-modulated two-dimensional semiconductor lateral heterostructures via layer-selected atomic substitution},
  author={Li, Honglai and Wu, Xueping and Liu, Hongjun and Zheng, Biyuan and Zhang, Qinglin and Zhu, Xiaoli and Wei, Zheng and Zhuang, Xiujuan and Zhou, Hong and Tang, Wenxin and others},
  journal={ACS nano},
  volume={11},
  number={1},
  pages={961--967},
  year={2017},
  publisher={ACS Publications}
}

@article{chen2015electronic,
  title={Electronic properties of MoS2--WS2 heterostructures synthesized with two-step lateral epitaxial strategy},
  author={Chen, Kun and Wan, Xi and Wen, Jinxiu and Xie, Weiguang and Kang, Zhiwen and Zeng, Xiaoliang and Chen, Huanjun and Xu, Jian-Bin},
  journal={ACS nano},
  volume={9},
  number={10},
  pages={9868--9876},
  year={2015},
  publisher={ACS Publications}
}

@article{zhang2015synthesis,
  title={Synthesis of lateral heterostructures of semiconducting atomic layers},
  author={Zhang, Xin-Quan and Lin, Chin-Hao and Tseng, Yu-Wen and Huang, Kuan-Hua and Lee, Yi-Hsien},
  journal={Nano letters},
  volume={15},
  number={1},
  pages={410--415},
  year={2015},
  publisher={ACS Publications}
}

@article{sahoo2018one,
  title={One-pot growth of two-dimensional lateral heterostructures via sequential edge-epitaxy},
  author={Sahoo, Prasana K and Memaran, Shahriar and Xin, Yan and Balicas, Luis and Guti{\'e}rrez, Humberto R},
  journal={Nature},
  volume={553},
  number={7686},
  pages={63--67},
  year={2018},
  publisher={Nature Publishing Group UK London}
}

@article{li2015epitaxial,
  title={Epitaxial growth of a monolayer WSe2-MoS2 lateral pn junction with an atomically sharp interface},
  author={Li, Ming-Yang and Shi, Yumeng and Cheng, Chia-Chin and Lu, Li-Syuan and Lin, Yung-Chang and Tang, Hao-Lin and Tsai, Meng-Lin and Chu, Chih-Wei and Wei, Kung-Hwa and He, Jr-Hau and others},
  journal={Science},
  volume={349},
  number={6247},
  pages={524--528},
  year={2015},
  publisher={American Association for the Advancement of Science}
}

@article{zhang2017robust,
  title={Robust epitaxial growth of two-dimensional heterostructures, multiheterostructures, and superlattices},
  author={Zhang, Zhengwei and Chen, Peng and Duan, Xidong and Zang, Ketao and Luo, Jun and Duan, Xiangfeng},
  journal={Science},
  volume={357},
  number={6353},
  pages={788--792},
  year={2017},
  publisher={American Association for the Advancement of Science}
}

@article{yang2017van,
  title={Van der Waals epitaxial growth and optoelectronics of large-scale WSe2/SnS2 vertical bilayer p--n junctions},
  author={Yang, Tiefeng and Zheng, Biyuan and Wang, Zhen and Xu, Tao and Pan, Chen and Zou, Juan and Zhang, Xuehong and Qi, Zhaoyang and Liu, Hongjun and Feng, Yexin and others},
  journal={Nature communications},
  volume={8},
  number={1},
  pages={1906},
  year={2017},
  publisher={Nature Publishing Group UK London}
}

@article{li2017two,
  title={Two-dimensional non-volatile programmable p--n junctions},
  author={Li, Dong and Chen, Mingyuan and Sun, Zhengzong and Yu, Peng and Liu, Zheng and Ajayan, Pulickel M and Zhang, Zengxing},
  journal={Nature nanotechnology},
  volume={12},
  number={9},
  pages={901--906},
  year={2017},
  publisher={Nature Publishing Group UK London}
}

@article{berweger2020spatially,
  title={Spatially resolved persistent photoconductivity in MoS2--WS2 lateral heterostructures},
  author={Berweger, Samuel and Zhang, Hanyu and Sahoo, Prasana K and Kupp, Benjamin M and Blackburn, Jeffrey L and Miller, Elisa M and Wallis, Thomas M and Voronine, Dmitri V and Kabos, Pavel and Nanayakkara, Sanjini U},
  journal={ACS nano},
  volume={14},
  number={10},
  pages={14080--14090},
  year={2020},
  publisher={ACS Publications}
}

@article{zhang2016visualizing,
  title={Visualizing band offsets and edge states in bilayer--monolayer transition metal dichalcogenides lateral heterojunction},
  author={Zhang, Chendong and Chen, Yuxuan and Huang, Jing-Kai and Wu, Xianxin and Li, Lain-Jong and Yao, Wang and Tersoff, Jerry and Shih, Chih-Kang},
  journal={Nature communications},
  volume={7},
  number={1},
  pages={10349},
  year={2016},
  publisher={Nature Publishing Group UK London}
}

@article{sahoo2025quasiparticle,
  title={Quasiparticle Gap Renormalization Driven by Internal and External Screening in a WS 2 Device},
  author={Sahoo, Chakradhar and ’t Veld, Yann in and Jones, Alfred JH and Jiang, Zhihao and Lupi, Greta and Majchrzak, Paulina E and Hsieh, Kimberly and Watanabe, Kenji and Taniguchi, Takashi and Hofmann, Philip and others},
  journal={Physical Review Letters},
  volume={135},
  number={5},
  pages={056401},
  year={2025},
  publisher={APS}
}

@article{zhou2013intrinsic,
  title={Intrinsic structural defects in monolayer molybdenum disulfide},
  author={Zhou, Wu and Zou, Xiaolong and Najmaei, Sina and Liu, Zheng and Shi, Yumeng and Kong, Jing and Lou, Jun and Ajayan, Pulickel M and Yakobson, Boris I and Idrobo, Juan-Carlos},
  journal={Nano letters},
  volume={13},
  number={6},
  pages={2615--2622},
  year={2013},
  publisher={ACS Publications}
}

@article{zheng2018band,
  title={Band alignment engineering in two-dimensional lateral heterostructures},
  author={Zheng, Biyuan and Ma, Chao and Li, Dong and Lan, Jianyue and Zhang, Zhe and Sun, Xingxia and Zheng, Weihao and Yang, Tiefeng and Zhu, Chenguang and Ouyang, Gang and others},
  journal={Journal of the American Chemical Society},
  volume={140},
  number={36},
  pages={11193--11197},
  year={2018},
  publisher={ACS Publications}
}

@article{nugera2022bandgap,
  title={Bandgap engineering in 2D lateral heterostructures of transition metal dichalcogenides via controlled alloying},
  author={Nugera, Florence A and Sahoo, Prasana K and Xin, Yan and Ambardar, Sharad and Voronine, Dmitri V and Kim, Un Jeong and Han, Yoojoong and Son, Hyungbin and Guti{\'e}rrez, Humberto R},
  journal={Small},
  volume={18},
  number={12},
  pages={2106600},
  year={2022},
  publisher={Wiley Online Library}
}

@article{liu2016van,
  title={Van der Waals heterostructures and devices},
  author={Liu, Yuan and Weiss, Nathan O and Duan, Xidong and Cheng, Hung-Chieh and Huang, Yu and Duan, Xiangfeng},
  journal={Nature Reviews Materials},
  volume={1},
  number={9},
  pages={1--17},
  year={2016},
  publisher={Nature Publishing Group}
}

@article{castellanos2022van,
  title={Van der Waals heterostructures},
  author={Castellanos-Gomez, Andres and Duan, Xiangfeng and Fei, Zhe and Gutierrez, Humberto Rodriguez and Huang, Yuan and Huang, Xinyu and Quereda, Jorge and Qian, Qi and Sutter, Eli and Sutter, Peter},
  journal={Nature Reviews Methods Primers},
  volume={2},
  number={1},
  pages={58},
  year={2022},
  publisher={Nature Publishing Group UK London}
}

@article{rivera2015observation,
  title={Observation of long-lived interlayer excitons in monolayer MoSe2--WSe2 heterostructures},
  author={Rivera, Pasqual and Schaibley, John R and Jones, Aaron M and Ross, Jason S and Wu, Sanfeng and Aivazian, Grant and Klement, Philip and Seyler, Kyle and Clark, Genevieve and Ghimire, Nirmal J and others},
  journal={Nature communications},
  volume={6},
  number={1},
  pages={6242},
  year={2015},
  publisher={Nature Publishing Group UK London}
}

@article{wang2018colloquium,
  title={Colloquium: Excitons in atomically thin transition metal dichalcogenides},
  author={Wang, Gang and Chernikov, Alexey and Glazov, Mikhail M and Heinz, Tony F and Marie, Xavier and Amand, Thierry and Urbaszek, Bernhard},
  journal={Reviews of Modern Physics},
  volume={90},
  number={2},
  pages={021001},
  year={2018},
  publisher={APS}
}

@article{feng2014growth,
  title={Growth of large-area 2D MoS2 (1-x) Se2x semiconductor alloys},
  author={Feng, Qingliang and Zhu, Yiming and Hong, Jinhua and Zhang, Mei and Duan, Wenjie and Mao, Nannan and Wu, Juanxia and Xu, Hua and Dong, Fengliang and Lin, Fang and others},
  journal={Advanced Materials},
  volume={26},
  number={17},
  pages={2648--2653},
  year={2014}
}

@article{duan2016synthesis,
  title={Synthesis of WS2 x Se2--2 x alloy nanosheets with composition-tunable electronic properties},
  author={Duan, Xidong and Wang, Chen and Fan, Zheng and Hao, Guolin and Kou, Liangzhi and Halim, Udayabagya and Li, Honglai and Wu, Xueping and Wang, Yicheng and Jiang, Jianhui and others},
  journal={Nano letters},
  volume={16},
  number={1},
  pages={264--269},
  year={2016},
  publisher={ACS Publications}
}

@article{caldwell2010technique,
  title={Technique for the dry transfer of epitaxial graphene onto arbitrary substrates},
  author={Caldwell, Joshua D and Anderson, Travis J and Culbertson, James C and Jernigan, Glenn G and Hobart, Karl D and Kub, Fritz J and Tadjer, Marko J and Tedesco, Joseph L and Hite, Jennifer K and Mastro, Michael A and others},
  journal={ACS nano},
  volume={4},
  number={2},
  pages={1108--1114},
  year={2010},
  publisher={ACS Publications}
}

@article{yi2015review,
  title={A review on mechanical exfoliation for the scalable production of graphene},
  author={Yi, Min and Shen, Zhigang},
  journal={Journal of Materials Chemistry A},
  volume={3},
  number={22},
  pages={11700--11715},
  year={2015},
  publisher={Royal Society of Chemistry}
}

@article{chu2018energy,
  title={Energy-resolved photoconductivity mapping in a monolayer--bilayer WSe2 lateral heterostructure},
  author={Chu, Zhaodong and Han, Ali and Lei, Chao and Lopatin, Sergei and Li, Peng and Wannlund, David and Wu, Di and Herrera, Kevin and Zhang, Xixiang and MacDonald, Allan H and others},
  journal={Nano letters},
  volume={18},
  number={11},
  pages={7200--7206},
  year={2018},
  publisher={ACS Publications}
}

@article{lv2022femtomolar,
  title={Femtomolar-level molecular sensing of monolayer tungsten diselenide induced by heteroatom doping with long-term stability},
  author={Lv, Qian and Tan, Junyang and Wang, Zhijie and Yu, Lingxiao and Liu, Bilu and Lin, Junhao and Li, Jia and Huang, Zheng-Hong and Kang, Feiyu and Lv, Ruitao},
  journal={Advanced Functional Materials},
  volume={32},
  number={34},
  pages={2200273},
  year={2022},
  publisher={Wiley Online Library}
}

@article{kang2018direct,
  title={Direct growth of doping controlled monolayer WSe 2 by selenium-phosphorus substitution},
  author={Kang, Won Tae and Lee, Il Min and Yun, Seok Joon and Song, Young Il and Kim, Kunnyun and Kim, Do-Hwan and Shin, Yong Seon and Lee, Kiyoung and Heo, Jinseong and Kim, Young-Min and others},
  journal={Nanoscale},
  volume={10},
  number={24},
  pages={11397--11402},
  year={2018},
  publisher={Royal Society of Chemistry}
}

@article{kastl2019effects,
  title={Effects of defects on band structure and excitons in WS2 revealed by nanoscale photoemission spectroscopy},
  author={Kastl, Christoph and Koch, Roland J and Chen, Christopher T and Eichhorn, Johanna and Ulstrup, S{\o}ren and Bostwick, Aaron and Jozwiak, Chris and Kuykendall, Tevye R and Borys, Nicholas J and Toma, Francesca M and others},
  journal={ACS nano},
  volume={13},
  number={2},
  pages={1284--1291},
  year={2019},
  publisher={ACS Publications}
}

@article{ulstrup2019nanoscale,
  title={Nanoscale mapping of quasiparticle band alignment},
  author={Ulstrup, S{\o}ren and Giusca, Cristina E and Miwa, Jill A and Sanders, Charlotte E and Browning, Alex and Dudin, Pavel and Cacho, Cephise and Kazakova, Olga and Gaskill, D Kurt and Myers-Ward, Rachael L and others},
  journal={Nature communications},
  volume={10},
  number={1},
  pages={3283},
  year={2019},
  publisher={Nature Publishing Group UK London}
}

@article{kresse1996efficiency,
  title={Efficiency of ab-initio total energy calculations for metals and semiconductors using a plane-wave basis set},
  author={Kresse, Georg and Furthm{\"u}ller, J{\"u}rgen},
  journal={Computational materials science},
  volume={6},
  number={1},
  pages={15--50},
  year={1996},
  publisher={Elsevier}
}

@article{kresse1994theory,
  title={Theory of the crystal structures of selenium and tellurium: the effect of generalized-gradient corrections to the local-density approximation},
  author={Kresse, G and Furthm{\"u}ller, J and Hafner, JJPRB},
  journal={Physical Review B},
  volume={50},
  number={18},
  pages={13181},
  year={1994},
  publisher={APS}
}

@article{kresse1999ultrasoft,
  title={From ultrasoft pseudopotentials to the projector augmented-wave method},
  author={Kresse, Georg and Joubert, Daniel},
  journal={Physical review B},
  volume={59},
  number={3},
  pages={1758},
  year={1999},
  publisher={APS}
}

@article{blochl1994projector,
  title={Projector augmented-wave method},
  author={Bl{\"o}chl, Peter E},
  journal={Physical review B},
  volume={50},
  number={24},
  pages={17953},
  year={1994},
  publisher={APS}
}

@article{perdew1996generalized,
  title={Generalized gradient approximation made simple},
  author={Perdew, John P and Burke, Kieron and Ernzerhof, Matthias},
  journal={Physical review letters},
  volume={77},
  number={18},
  pages={3865},
  year={1996},
  publisher={APS}
}

@article{monkhorst1976special,
  title={Special points for Brillouin-zone integrations},
  author={Monkhorst, Hendrik J and Pack, James D},
  journal={Physical review B},
  volume={13},
  number={12},
  pages={5188},
  year={1976},
  publisher={APS}
}

@article{schutte1987crystal,
  title={Crystal structures of tungsten disulfide and diselenide},
  author={Schutte, WJ and De Boer, JL and Jellinek, F},
  journal={Journal of Solid State Chemistry},
  volume={70},
  number={2},
  pages={207--209},
  year={1987},
  publisher={Elsevier}
}

@article{zhang2014direct,
  title={Direct observation of the transition from indirect to direct bandgap in atomically thin epitaxial MoSe2},
  author={Zhang, Yi and Chang, Tay-Rong and Zhou, Bo and Cui, Yong-Tao and Yan, Hao and Liu, Zhongkai and Schmitt, Felix and Lee, James and Moore, Rob and Chen, Yulin and others},
  journal={Nature nanotechnology},
  volume={9},
  number={2},
  pages={111--115},
  year={2014},
  publisher={Nature Publishing Group UK London}
}

@article{desai2014strain,
  title={Strain-induced indirect to direct bandgap transition in multilayer WSe2},
  author={Desai, Sujay B and Seol, Gyungseon and Kang, Jeong Seuk and Fang, Hui and Battaglia, Corsin and Kapadia, Rehan and Ager, Joel W and Guo, Jing and Javey, Ali},
  journal={Nano letters},
  volume={14},
  number={8},
  pages={4592--4597},
  year={2014},
  publisher={ACS Publications}
}

@article{lu2017identifying,
  title={Identifying and visualizing the edge terminations of single-layer MoSe2 island epitaxially grown on Au (111)},
  author={Lu, Jianchen and Bao, De-Liang and Qian, Kai and Zhang, Shuai and Chen, Hui and Lin, Xiao and Du, Shi-Xuan and Gao, Hong-Jun},
  journal={ACS nano},
  volume={11},
  number={2},
  pages={1689--1695},
  year={2017},
  publisher={ACS Publications}
}

@article{haldar2015systematic,
  title={Systematic study of structural, electronic, and optical properties of atomic-scale defects in the two-dimensional transition metal dichalcogenides MX 2 (M= Mo, W; X= S, Se, Te)},
  author={Haldar, Soumyajyoti and Vovusha, Hakkim and Yadav, Manoj Kumar and Eriksson, Olle and Sanyal, Biplab},
  journal={Physical Review B},
  volume={92},
  number={23},
  pages={235408},
  year={2015},
  publisher={APS}
}

@article{shaposhnikov2019impact,
  title={Impact of defects on electronic properties of heterostructures constructed from monolayers of transition metal dichalcogenides},
  author={Shaposhnikov, Victor L and Krivosheeva, Anna V and Borisenko, Victor E},
  journal={physica status solidi (b)},
  volume={256},
  number={5},
  pages={1800355},
  year={2019},
  publisher={Wiley Online Library}
}

@article{kucinski2024direct,
  title={Direct Measurement of the Thermal Expansion Coefficient of Epitaxial WSe2 by Four-Dimensional Scanning Transmission Electron Microscopy},
  author={Kucinski, Theresa M and Dhall, Rohan and Savitzky, Benjamin H and Ophus, Colin and Karkee, Rijan and Mishra, Avanish and Dervishi, Enkeleda and Kang, Jung Hoon and Lee, Chul-Ho and Yoo, Jinkyoung and others},
  journal={ACS nano},
  volume={18},
  number={27},
  pages={17725--17734},
  year={2024},
  publisher={ACS Publications}
}

@article{zhong2022unified,
  title={A unified approach and descriptor for the thermal expansion of two-dimensional transition metal dichalcogenide monolayers},
  author={Zhong, Yang and Zhang, Lenan and Park, Ji-Hoon and Cruz, Samuel and Li, Long and Guo, Liang and Kong, Jing and Wang, Evelyn N},
  journal={Science Advances},
  volume={8},
  number={46},
  pages={eabo3783},
  year={2022},
  publisher={American Association for the Advancement of Science}
}

@article{komsa2012two,
  title={Two-Dimensional Transition Metal Dichalcogenides under Electron Irradiation:<? format?> Defect Production and Doping},
  author={Komsa, Hannu-Pekka and Kotakoski, Jani and Kurasch, Simon and Lehtinen, Ossi and Kaiser, Ute and Krasheninnikov, Arkady V},
  journal={Physical review letters},
  volume={109},
  number={3},
  pages={035503},
  year={2012},
  publisher={APS}
}

@article{komsa2015native,
  title={Native defects in bulk and monolayer MoS 2 from first principles},
  author={Komsa, Hannu-Pekka and Krasheninnikov, Arkady V},
  journal={Physical Review B},
  volume={91},
  number={12},
  pages={125304},
  year={2015},
  publisher={APS}
}

@article{han2018strain,
  title={Strain mapping of two-dimensional heterostructures with subpicometer precision},
  author={Han, Yimo and Nguyen, Kayla and Cao, Michael and Cueva, Paul and Xie, Saien and Tate, Mark W and Purohit, Prafull and Gruner, Sol M and Park, Jiwoong and Muller, David A},
  journal={Nano letters},
  volume={18},
  number={6},
  pages={3746--3751},
  year={2018},
  publisher={ACS Publications}
}

@article{bogaert2016diffusion,
  title={Diffusion-mediated synthesis of MoS2/WS2 lateral heterostructures},
  author={Bogaert, Kevin and Liu, Song and Chesin, Jordan and Titow, Denis and Gradecak, Silvija and Garaj, Slaven},
  journal={Nano letters},
  volume={16},
  number={8},
  pages={5129--5134},
  year={2016},
  publisher={ACS Publications}
}

@article{taghinejad2018defect,
  title={Defect-mediated alloying of monolayer transition-metal dichalcogenides},
  author={Taghinejad, Hossein and Rehn, Daniel A and Muccianti, Christine and Eftekhar, Ali A and Tian, Mengkun and Fan, Tianren and Zhang, Xiang and Meng, Yuze and Chen, Yanwen and Nguyen, Tran-Vinh and others},
  journal={ACS nano},
  volume={12},
  number={12},
  pages={12795--12804},
  year={2018},
  publisher={ACS Publications}
}

@article{holtzman2024equilibrium,
  title={Equilibrium densities of intrinsic defects in transition metal diselenides of molybdenum and tungsten},
  author={Holtzman, Luke N and Vargas, Preston Allen and Hennig, Richard G and Barmak, Katayun},
  journal={The Journal of Chemical Physics},
  volume={161},
  number={14},
  year={2024},
  publisher={AIP Publishing}
}

@article{miwa2015van,
  title={Van der Waals epitaxy of two-dimensional MoS2--graphene heterostructures in ultrahigh vacuum},
  author={Miwa, Jill A and Dendzik, Maciej and Gr{\o}nborg, Signe S and Bianchi, Marco and Lauritsen, Jeppe V and Hofmann, Philip and Ulstrup, S{\o}ren},
  journal={ACS nano},
  volume={9},
  number={6},
  pages={6502--6510},
  year={2015},
  publisher={ACS Publications}
}

@article{madeo2020directly,
  title={Directly visualizing the momentum-forbidden dark excitons and their dynamics in atomically thin semiconductors},
  author={Mad{\'e}o, Julien and Man, Michael KL and Sahoo, Chakradhar and Campbell, Marshall and Pareek, Vivek and Wong, E Laine and Al-Mahboob, Abdullah and Chan, Nicholas S and Karmakar, Arka and Mariserla, Bala Murali Krishna and others},
  journal={Science},
  volume={370},
  number={6521},
  pages={1199--1204},
  year={2020},
  publisher={American Association for the Advancement of Science}
}

@article{stansbury2021visualizing,
  title={Visualizing electron localization of WS2/WSe2 moir{\'e} superlattices in momentum space},
  author={Stansbury, Conrad H and Utama, M Iqbal Bakti and Fatuzzo, Claudia G and Regan, Emma C and Wang, Danqing and Xiang, Ziyu and Ding, Mingchao and Watanabe, Kenji and Taniguchi, Takashi and Blei, Mark and others},
  journal={Science advances},
  volume={7},
  number={37},
  pages={eabf4387},
  year={2021},
  publisher={American Association for the Advancement of Science}
}

@article{nakamura2020spin,
  title={Spin splitting and strain in epitaxial monolayer WSe 2 on graphene},
  author={Nakamura, H and Mohammed, A and Rosenzweig, Ph and Matsuda, K and Nowakowski, K and K{\"u}ster, K and Wochner, P and Ibrahimkutty, S and Wedig, U and Hussain, H and others},
  journal={Physical Review B},
  volume={101},
  number={16},
  pages={165103},
  year={2020},
  publisher={APS}
}

@article{gatti2023flat,
  title={Flat $\Gamma$ moir{\'e} bands in twisted bilayer WSe 2},
  author={Gatti, Gianmarco and Issing, Julia and Rademaker, Louk and Margot, Florian and de Jong, Tobias A and van der Molen, Sense Jan and Teyssier, J{\'e}r{\'e}mie and Kim, Timur K and Watson, Matthew D and Cacho, Cephise and others},
  journal={Physical Review Letters},
  volume={131},
  number={4},
  pages={046401},
  year={2023},
  publisher={APS}
}

@article{karni2022structure,
  title={Structure of the moir{\'e} exciton captured by imaging its electron and hole},
  author={Karni, Ouri and Barr{\'e}, Elyse and Pareek, Vivek and Georgaras, Johnathan D and Man, Michael KL and Sahoo, Chakradhar and Bacon, David R and Zhu, Xing and Ribeiro, Henrique B and O’Beirne, Aidan L and others},
  journal={Nature},
  volume={603},
  number={7900},
  pages={247--252},
  year={2022},
  publisher={Nature Publishing Group UK London}
}

@article{rosati2023interface,
  title={Interface engineering of charge-transfer excitons in 2D lateral heterostructures},
  author={Rosati, Roberto and Paradisanos, Ioannis and Huang, Libai and Gan, Ziyang and George, Antony and Watanabe, Kenji and Taniguchi, Takashi and Lombez, Laurent and Renucci, Pierre and Turchanin, Andrey and others},
  journal={Nature communications},
  volume={14},
  number={1},
  pages={2438},
  year={2023},
  publisher={Nature Publishing Group UK London}
}

@article{vashishtha2025epitaxial,
  title={Epitaxial Interface-Driven Photoresponse Enhancement in Monolayer WS2--MoS2 Lateral Heterostructures},
  author={Vashishtha, Pargam and Kofler, Clara and Verma, Ajay Kumar and Giridhar, Sindhu Priya and Tollerud, Jonathan O and Dissanayake, Nethmi SL and Gupta, Tanish and Sehrawat, Manoj and Aggarwal, Vishnu and Mayes, Edwin LH and others},
  journal={Advanced Functional Materials},
  pages={e12962},
  year={2025},
  publisher={Wiley Online Library}
}

@article{yao2017optically,
  title={Optically discriminating carrier-induced quasiparticle band gap and exciton energy renormalization in monolayer MoS 2},
  author={Yao, Kaiyuan and Yan, Aiming and Kahn, Salman and Suslu, Aslihan and Liang, Yufeng and Barnard, Edward S and Tongay, Sefaattin and Zettl, Alex and Borys, Nicholas J and Schuck, P James},
  journal={Physical Review Letters},
  volume={119},
  number={8},
  pages={087401},
  year={2017},
  publisher={APS}
}

@misc{SahooLHSSI,
	note = {{See Supplemental Material for LHS growth details, for details of EDC analysis of valence band states, for extraction of VBM and exciton mean energies, for core level, for details of $DFT$ calculations.}}}

@article{sadhukhan2022generating,
  title={Generating bright emissive states by modulating the bandgap of monolayer tungsten diselenide},
  author={Sadhukhan, Tumpa and Schatz, George C},
  journal={The Journal of Physical Chemistry C},
  volume={126},
  number={12},
  pages={5598--5606},
  year={2022},
  publisher={ACS Publications}
}
 \end{document}